\documentclass[aps]{revtex4}%
\usepackage{amsfonts}
\usepackage{amsmath}
\usepackage{amssymb}
\usepackage{graphicx}%
\begin{document}
\title[Graphene resistivity]{Short-range defects contribution to the monolayer graphene resistivity}
\author{Natalie E. Firsova}
\affiliation{Institute for Problems of Mechanical Engineering, the Russian Academy of
Sciences, St. Petersburg 199178, Russia}
\author{Sergey A. Ktitorov}
\affiliation{A.F. Ioffe Physical-Technical Institute, the Russian Academy of Sciences, St.
Petersburg, Russia}
\author{}

\begin{abstract}
The derived by us earlier electron scattering matrix for the short-range
defects in monolayer graphene is applyed to description of the resistivity
electron density dependence. It is argued that large charged defect density is
unlikely in the suspended graphene and graphene lying above the trench. 

\end{abstract}
\maketitle








\section{Introduction}

Electronic states in monolayer graphene with point defects were considered in
\cite{basko}, \cite{pogorel}, \cite{novic} and in our works \cite{we1},
\cite{we2}. The main novel element in our works is the band asymmetry of the
defect potential in the Dirac equation. This asymmetry appears naturally if
the defect violates the symmetry between the sublattices. This asymmetry was
described in terms of equivalent superposition the band-symmetric potential
and the mass (gap) perturbations. Characteristic equations for the energy
eigenvalues for the gapped graphene and complex energies for resonances were
derived and analysed. Exact formulae for the scattering and transfer matrices
were obtained. Important features of the electronic states in the presence of
the asymmetric potential were found in \ref{we2}. Our goal in this work is to
study the impact of these features on the electronic transport in the
monolayer graphene.

\section{Basic equations}

The Dirac equation describing electronic states in zero-gap graphene reads
\cite{novosel}
\begin{equation}
\left(  -iv_{F}\hbar\sum_{\mu=1}^{2}\sigma_{\mu}\partial_{\mu}-\sigma
_{3}\delta mv_{F}^{{}^{2}}\right)  \psi=\left(  E-V\right)  \psi,
\label{diracgeneral}%
\end{equation}
where $v_{F}$ is the Fermi velocity, $\sigma_{\mu}$ are the Pauli matrices,
$\psi\left(  \mathbf{r}\right)  $ is the two-component spinor.The spinor
structure takes into account the two-sublattice structure of graphene
$\ \delta m\left(  \mathbf{r}\right)  $ and $V(\mathbf{r})$ are the local
perturbations of the mass (gap) and the chemical potential. A local mass
perturbation can be induced by defects in the graphene film or in the
substrate. The perturbation matrix elements
\begin{equation}
diag(V_{1},V_{2})r_{0}\delta(r-r_{0}) \label{diag}%
\end{equation}
are related to the $a,$ $b$ parameters as follows
\begin{equation}
-V_{1}=a+b,\text{ }-V_{2}=a-b \label{abVrelation}%
\end{equation}

The delta function perturbation is the simplest solvable short-range model.
Finite radius $r_{0}$ plays a role of the regulator and is necessary in order
to exclude deep states of the atomic energy scale. The finite perturbation
radius $r_{0}$ leads to the quasi-momentum space form-factor proportional to
the Bessel function that justifies our neglect of transitions between the
Brillouin band points $K$\ and $K^{\prime}$.

Let us introduce dimensionalless variables and parameters:%
\begin{align}
\epsilon &  =\frac{E}{\hbar v_{F}/r_{0}},\text{ \ \ }\widetilde{m}%
=\frac{mv_{F}{}^{2}}{\hbar v_{F}/r_{0}},\text{ \ }u_{i}=\frac{V_{i}}{\hbar
v_{F}/r_{0}},\text{ \ }i=1,\text{ }2,\text{ \ }\widetilde{\mathbf{r}}%
=\frac{\mathbf{r}}{r_{0}},\text{ \ }\widetilde{\partial}_{_{\mu}}%
=\frac{\partial}{\partial\widetilde{\mathbf{r}}},\label{dimless}\\
r_{0}\delta(r-r_{0})  &  =\delta(\widetilde{r}-1).
\end{align}

The equation (\ref{diracgeneral}) takes the form%
\begin{equation}
\left(  -i\sum_{\mu=1}^{2}\sigma_{\mu}\widetilde{\partial}_{\mu}+\sigma
_{3}\widetilde{m}-\sigma_{3}\delta\widetilde{m}\right)  \psi=\left(
\epsilon-u\right)  \psi, \label{diracdimless}%
\end{equation}

We confine ourselves here by the case of zero-gap graphene; $\widetilde{m}=0.$

Let us present the two-component spinor in the form%
\begin{equation}
\psi_{j}(\widetilde{\mathbf{r}},t)=\frac{\exp\left(  -iEt\right)  }%
{\sqrt{\widetilde{r}}}%
\begin{pmatrix}
f_{j}\left(  \widetilde{r}\right)  \exp\left[  i\left(  j-1/2\right)
\phi\right] \\
g_{j}\left(  \widetilde{r}\right)  \exp\left[  i\left(  j+1/2\right)
\phi\right]
\end{pmatrix}
, \label{spinor}%
\end{equation}
where $j$ is the pseudospin quantum number; $j=\pm1/2,$ $\pm3/2,\ldots$. In
the opposite to the relativistic theory, this quantum number has nothing to do
with the real spin and indicates the degeneracy in the biconic Dirac point.
The upper $f_{j}\left(  r\right)  $ and $\ $lower $g_{j}\left(  r\right)  $
components of the spinor satisfy the equations
\begin{equation}
\frac{dg_{j}}{d\widetilde{r}}+\frac{j}{\widetilde{r}}g_{j}+\epsilon
f_{j}=\left(  a+b\right)  \delta(\widetilde{r}-1)f_{j}, \label{componenteq1}%
\end{equation}

\begin{equation}
-\frac{df_{j}}{d\widetilde{r}}+\frac{j}{\widetilde{r}}f_{j}-\epsilon
g_{j}=\left(  a-b\right)  \delta(\widetilde{r}-1)g_{j}. \label{componeq2}%
\end{equation}

The general solution can be found solving the second-order equation obtained
by excluding one of the spinor components from the equation set
(\ref{componenteq1}), (\ref{componeq2}) in the domains $0<r<r_{0}$ and
$r>r_{0}:$%
\begin{equation}
\frac{d^{2}f_{j}}{d\widetilde{r}^{2}}+\left[  \epsilon^{2}-\frac{j\left(
j-1\right)  }{\widetilde{r}^{2}}\right]  f_{j}=0. \label{secondorder}%
\end{equation}
This equation is related to the Bessel one. Its general solution in the domain
$0<r<r_{0}$ reads%

\begin{equation}
f_{j}=C_{1}\sqrt{\epsilon\widetilde{r}}J_{j-1/2}\left(  \epsilon\widetilde
{r}\right)  +C_{2}\sqrt{\epsilon\widetilde{r}}N_{j-1/2}\left(  \epsilon
\widetilde{r}\right)  , \label{general}%
\end{equation}

where $J_{\nu}\left(  z\right)  $ and $N_{\nu}\left(  z\right)  $ are the
Bessel and Neumann functions respectively.The constant $C_{2\text{ }}$vanishes
in the domain $0\leq\widetilde{r}<1$ since the solution must be regular at the
origin. The solution in the region $\widetilde{r}\succeq1$ reads%

\[
f_{j}=C_{3}H_{j-1/2}^{\left(  1\right)  }\left(  \epsilon\widetilde{r}\right)
+C_{4}H_{j-1/2}^{\left(  1\right)  }\left(  \epsilon\widetilde{r}\right)  ,
\]

where $H_{\nu}^{\left(  \alpha\right)  }\left(  z\right)  $ is Hankel's
function. Matching these solutions at the circumference of the circle of
radius $r=r_{0}$ $\left(  \widetilde{r}=1\right)  $ we obtain a scattering
matrix and a characteristic equation for the resonance states. Calculating the
ratio of the out-going and in-going waves, we obtain the S-matrix components
in the angular momentum representation (for details see \cite{we2}):%

\begin{equation}
S_{j}\left(  \epsilon\right)  =-\frac{\mathcal{F}_{j}^{\left(  2\right)  }%
}{\mathcal{F}_{j}^{\left(  1\right)  }}, \label{S2}%
\end{equation}

where $\mathcal{F}_{j}^{\left(  \alpha\right)  }$ is given by the formula:%
\begin{align}
\mathcal{F}_{j}^{\left(  \alpha\right)  }  &  =\left(  J_{j-1/2}\left(
\epsilon\right)  H_{j+1/2}^{\left(  \alpha\right)  }\left(  \epsilon\right)
-J_{j+1/2}\left(  \epsilon\right)  H_{j-1/2}^{\left(  \alpha\right)  }\left(
\epsilon\right)  \right)  -\nonumber\\
&  \left[  \left(  a-b\right)  J_{j+1/2}\left(  \epsilon\right)
H_{j+1/2}^{\left(  \alpha\right)  }\left(  \epsilon\right)  +\left(
a+b\right)  J_{j-1/2}\left(  \epsilon\right)  H_{j-1/2}^{\left(
\alpha\right)  }\left(  \epsilon\right)  \right]  ,\nonumber\\
\alpha &  =1,\text{ \ }2. \label{F}%
\end{align}

Poles of the scattering matrix (\ref{S2}) are determined by the characteristic equation%

\begin{equation}
\mathcal{F}_{j}^{\left(  1\right)  }\left(  \epsilon\right)  =0, \label{char}%
\end{equation}

or%

\begin{equation}
\left(  a-b\right)  J_{j+1/2}^{2}\left(  \epsilon\right)  +\left(  a+b\right)
J_{j-1/2}^{2}\left(  \epsilon\right)  =i\left[  \left(  a-b\right)
J_{j+1/2}\left(  \epsilon\right)  N_{j+1/2}\left(  \epsilon\right)  +\left(
a+b\right)  J_{j-1/2}\left(  \epsilon\right)  N_{j-1/2}\left(  \epsilon
\right)  \right]  =-i\frac{2}{\pi\epsilon} \label{char2}%
\end{equation}

\section{Analysis of the characteristic equation and calculation of the
conductivity}

Using the relations $H_{n}^{\left(  1\right)  }\left(  z\right)  =J_{n}%
+iN_{n},$ $H_{n}^{\left(  2\right)  }=J_{n}-iN_{n},$ we can write S-matrix in
the form:%
\begin{equation}
S_{j}\left(  \epsilon\right)  =-\frac{A_{j}\left(  \epsilon\right)
-iB_{j}\left(  \epsilon\right)  }{A_{j}\left(  \epsilon\right)  +iB_{j}\left(
\epsilon\right)  }=\frac{B_{j}\left(  \epsilon\right)  +iA_{j}\left(
\epsilon\right)  }{B_{j}\left(  \epsilon\right)  -iA_{j}\left(  \epsilon
\right)  }, \label{swrational}%
\end{equation}
and, therefore, it can be presented in the standard form \cite{KMLL}%
\begin{equation}
S_{j}\left(  \epsilon\right)  =\exp\left[  i2\delta_{j}\left(  \epsilon
\right)  \right]  , \label{phase}%
\end{equation}
where the scattering phase is given by the expression
\begin{equation}
\delta_{j}\left(  \epsilon\right)  =\arctan\frac{A_{j}\left(  \epsilon\right)
}{B_{j}\left(  \epsilon\right)  }. \label{delta}%
\end{equation}
Formulae (\ref{swrational}), (\ref{phase}) show that the scattering matrix
$S_{j}\left(  \epsilon\right)  $ is unitary on the continuum spectrum. The
functions $A_{j}\left(  \epsilon\right)  $ and $B_{j}\left(  \epsilon\right)
$ are determined as follows%
\begin{equation}
A_{j}\left(  \epsilon\right)  =-\left[  \left(  a+b\right)  J_{j-1/2}%
^{2}\left(  \epsilon\right)  +\left(  a-b\right)  J_{j+1/2}^{2}\left(
\epsilon\right)  \right]  , \label{A}%
\end{equation}

\begin{align}
B_{j}\left(  \epsilon\right)   &  =-\frac{2}{\pi\epsilon}\left[  \left(
a+b\right)  J_{j-1/2}\left(  \epsilon\right)  N_{j-1/2}\left(  \epsilon
\right)  +\left(  a-b\right)  J_{j+1/2}\left(  \epsilon\right)  N_{j+1/2}%
\left(  \epsilon\right)  \right]  +\nonumber\\
&  \left[  J_{j+1/2}\left(  \epsilon\right)  N_{j-1/2}\left(  \epsilon\right)
-J_{j-1/2}\left(  \epsilon\right)  N_{j+1/2}\left(  \epsilon\right)  \right]
\label{B}%
\end{align}
Asymptotic behaviour of the scattering phases and other scattering data at
$\epsilon\rightarrow0$ can be obtained expanding the cylinder functions for
small arguments \cite{stegun}:
\begin{equation}
J_{n}\left(  z\right)  \sim\frac{1}{n!}\left(  z/2\right)  ^{2},\text{
\ \ }N_{n}\left(  z\right)  =\left\{
\begin{array}
[c]{c}%
-\frac{\Gamma\left(  n\right)  }{\pi}\left(  2/z\right)  ^{n}\text{ \ \ for
}n>0,\\
\left(  2/\pi\right)  \log\left(  \gamma_{E}z/2\right)  \text{ \ \ for
}n=0,\text{ \ \ }z\rightarrow0,
\end{array}
\right.  \label{expansion}%
\end{equation}
where $\gamma_{E}\approx0,577$ is the Eyler-Masceroni constant, $\Gamma\left(
n\right)  $ is the gamma-function. Then we have for the scattering phases in
the lower order in $\epsilon$:%
\begin{equation}
\tan\delta_{\pm1/2}\approx\left(  b\pm a\right)  \frac{\pi}{2}\epsilon
,\text{\ \ \ }\epsilon\rightarrow0, \label{phaseexp}%
\end{equation}

\begin{equation}
\tan\delta_{\pm\left(  n+1/2\right)  }\approx\pm\pi\left(  \epsilon/2\right)
^{2n+1}\left(  b\pm a\right)  \text{ \ \ }\epsilon\rightarrow0.
\label{phaseexp2}%
\end{equation}

The transport cross section can be written in terms of the scattering phases
\cite{novic} (we have returned to dimensional variables here)%
\begin{equation}
\Xi_{tr}=\frac{2r_{0}}{\epsilon}\sum_{j=\pm\frac{1}{2},\pm\frac{3}{2},\ldots
}\sin^{2}\left(  \delta_{j+1}-\delta_{j}\right)  . \label{cross}%
\end{equation}
The transport relaxation time can be calculated using the following relation:%
\begin{equation}
1/\tau_{tr}=N_{I}v_{F}\Xi_{tr}. \label{time}%
\end{equation}

Taking into account the approximate formulae for phases (\ref{phaseexp}),
(\ref{phaseexp2}), the series (\ref{cross}) can be written in the following
asymptotic form for $\epsilon\rightarrow0$:%

\begin{align}
\Xi_{tr}  &  =\frac{2r_{0}}{\epsilon}\left[  \left(  \delta_{1/2}%
-\delta_{-1/2}\right)  ^{2}+\left(  \delta_{3/2}-\delta_{1/2}\right)
^{2}+\left(  \delta_{-3/2}-\delta_{-1/2}\right)  ^{2}+\cdots\right.
\nonumber\\
\left.  \left(  \delta_{n+1/2}-\delta_{n-1/2}\right)  ^{2}+\left(
\delta_{-n-1/2}-\delta_{-n+1/2}\right)  ^{2}+\cdots\right]   &  \approx
\epsilon r_{0}\pi^{2}\left[  2a^{2}+O\left(  \epsilon\right)  \right]  .
\label{cross2}%
\end{align}

Then the asymptotic formula for the transport relaxation time reads:%
\begin{equation}
1/\tau_{tr}=\epsilon N_{I}v_{F}\pi^{2}2a^{2}\left[  1+O\left(  \epsilon
\right)  \right]  \label{time2}%
\end{equation}

It is seen from (\ref{time2}) that asymptotic behaviour of the relaxation time
at $\epsilon\rightarrow0$ is determined by the parameter $a,$ i. e. by the
symmetric component of the perturbation.

Let us consider now the Born approximation for the scattering amplitude. The
partial wave series for the transport cross section converges rather slowly.
That is why we consider a behaviour of the transport cross section without the
partial wave expansion. In return we can use the Born approximation in this
limit. The Born formula for the scattering amplitude reads \cite{novic}:%
\begin{equation}
f^{Born}\left(  p,\theta\right)  =-\frac{1}{\hbar v_{F}}\sqrt{\frac{p}{8\pi}%
}V\left(  \mathbf{q}\right)  , \label{amplitude}%
\end{equation}
where $\hbar\mathbf{q=\hbar p-\hbar p}^{\prime}$ is the transferred momentum,
$q=2p\sin\theta/2,$ $V\left(  \mathbf{q}\right)  $ is the perturbation Fourier transform:%

\begin{equation}
V\left(  \mathbf{q}\right)  =\int d^{2}re^{-i\mathbf{qr}}V\left(
\mathbf{r}\right)  =\int_{0}^{\infty}drrV\left(  r\right)  \int_{0}^{2\pi
}d\phi\exp\left[  -iqr\cos\phi\right]  =2\pi\int_{0}^{\infty}drrV\left(
r\right)  J_{0}\left(  qr\right)  . \label{fourierpotential}%
\end{equation}

Inserting the potental (\ref{diag}) into (\ref{fourierpotential}) we obtain
\begin{equation}
V_{i}\left(  \mathbf{q}\right)  \equiv V\left(  p,\theta\right)  =2\pi
V_{i}^{0}r_{0}^{2}J_{0}\left(  2pr_{0}\sin\theta/2\right)  .
\label{fourierpot2}%
\end{equation}

Substituting (\ref{fourierpot2}) into (\ref{amplitude}) we obtain the
scattering amplitude:%
\begin{equation}
f_{i}^{Born}\left(  p,\theta\right)  =-\frac{2\pi r_{0}^{2}V_{i}^{0}}{\hbar
v_{F}}\sqrt{\frac{p}{8\pi}}J_{0}\left(  2pr_{0}\sin\theta/2\right)  ,
\label{amplitude2}%
\end{equation}

Now we can calculate the transport cross section \cite{novic}:%

\begin{equation}
\Xi_{tr}^{Born}=\int_{0}^{\pi}d\theta\left(  1-\cos\theta\right)  \left\vert
f^{Born}\left(  p,\theta\right)  \right\vert ^{2}=\left(  pr_{0}\right)
r_{0}\left(  \frac{V_{i}^{0}}{\hbar v_{F}/r_{0}}\right)  ^{2}\pi/2\int
_{0}^{\pi}d\theta\left(  1-\cos\theta\right)  J_{0}^{2}\left(  2pr_{0}%
\sin\theta/2\right)  . \label{crossborn}%
\end{equation}

This integral can be expressed in terms of the hypergeometric functions
\cite{stegun}:%

\begin{align}
\int_{0}^{\pi}d\theta\left(  1-\cos\theta\right)  J_{0}^{2}\left(  2pr_{0}%
\sin\theta/2\right)   &  =\Gamma\left(
\begin{array}
[c]{ccc}%
1/2, & 3/2 & \\
2, & 1, & 1
\end{array}
\right)  \cdot\text{ }\nonumber\\
&  _{3}F_{4}\left(  3/2,1/2,1,2,1,1,1;-\left(  2pr_{0}\right)  ^{2}\right)  ,
\label{hyper}%
\end{align}
where $\Gamma\left(
\begin{array}
[c]{ccc}%
\alpha_{1}, & \alpha_{2} & \\
\beta_{1}, & \beta_{2}, & \beta_{3}%
\end{array}
\right)  \equiv\frac{\Gamma\left(  \alpha_{1}\right)  \Gamma\left(  \alpha
_{2}\right)  }{\Gamma\left(  \beta_{1}\right)  \Gamma\left(  \beta_{2}\right)
\Gamma\left(  \beta_{3}\right)  },$ $\Gamma\left(  \alpha\right)  $ is the
gamma function, $_{3}F_{4}\left(  \alpha_{1},\alpha_{2},\alpha_{3};\beta
_{1},\beta_{2},\beta_{3},\beta_{4};x\right)  $ is the generalized
hypergeometric function. It is determined by the series \cite{stegun}:%
\begin{equation}
_{3}F_{4}\left(  \alpha_{1},\alpha_{2},\alpha_{3};\beta_{1},\beta_{2}%
,\beta_{3},\beta_{4};x\right)  =\sum_{k=0}^{\infty}\frac{\left(  \alpha
_{1}\right)  _{k}\left(  \alpha_{2}\right)  _{k}\left(  \alpha_{3}\right)
_{k}}{\left(  \beta_{1}\right)  _{k}\left(  \beta_{2}\right)  _{k}\left(
\beta_{3}\right)  _{k}\left(  \beta_{4}\right)  _{k}}\frac{x^{k}}{k!},
\label{hyperseries}%
\end{equation}
where $\left(  \alpha\right)  _{k}=\frac{\Gamma\left(  \alpha+k\right)
}{\Gamma\left(  k\right)  }$ is the rising Pohhammer symbol. When $pr_{0}<1,$
we can neglect all higher terms of this series and otain in result the
transport scattering cross section $\Xi_{tr}^{Born}\symbol{126}pr_{0}.$ Notice
that the Born approximation is here asymptotically exact in the limit
$pr_{0}\rightarrow0.$ It must be noted here that the limit of $pr_{0},$
$E_{F},$ $k_{B}T,$ $\hbar\omega$ tending to zero is obviously nontrivial and
many-particle effects must be taken into account in this case. This problem
will be not discussed here. The opposite limit of $pr_{0}>1$ is not actual for
scattering on rare point defects. It can be urgent for the model of random
potential with large correlation radius and for the quantum dot model, but
they are outside of the scope of this work. However, there exist at least two
other possibilities to obtan the cross section behaviour, different from the
power law $\Xi_{tr}^{Born}\symbol{126}pr_{0}.$ First of them corresponds to a
random potential with the correlation radius $r_{0}>>a.$ In this case one can
consider scattering of electrons with $pr_{0}\sim1.$ Another possibility is
related to the resonance scattering studied in our works \cite{we2}. We
analyse here the latter case.

Let us consider the limit of large angular momentum $j>>1$ using the Bessel
function asymptotics:%
\begin{equation}
J_{\nu}\left(  z\right)  \sim\frac{1}{\sqrt{2\pi\nu}}\left(  \frac{ez}{2\nu
}\right)  ^{\nu}\left(  1+O\left(  1/\nu\right)  \right)  ,\text{ \ \ }%
\nu\rightarrow\infty, \label{largejJ}%
\end{equation}

\begin{equation}
N_{\nu}\left(  z\right)  \sim-\sqrt{\frac{2}{\pi\nu}}\left(  \frac{ez}{2\nu
}\right)  ^{-\nu}\left(  1+O\left(  1/\nu\right)  \right)  ,\text{ \ \ }%
\nu\rightarrow\infty, \label{largejN}%
\end{equation}

\begin{equation}
H_{\nu}^{\left(  1\right)  }\left(  z\right)  \sim\frac{1}{\sqrt{2\pi\nu}%
}\left[  \left(  \frac{ez}{2\nu}\right)  ^{\nu}-2i\left(  \frac{ez}{2\nu
}\right)  ^{-\nu}\right]  \left(  1+O\left(  1/\nu\right)  \right)  ,\text{
\ \ }\nu\rightarrow\infty, \label{largejH}%
\end{equation}
where $e$ is the base of natural logarithms. The characteristic equation
\ref{char2} takes the following form in the limit of $j>>1:$%

\begin{equation}
\frac{a-b}{\left(  2j+1\right)  \left(  2j-1\right)  ^{2j+2}}\left(
e\epsilon\right)  ^{2j+2}+\frac{a+b}{\left(  2j-1\right)  \left(  2j-1\right)
^{2j}}\left(  e\epsilon\right)  ^{2j}=ie\left[  \frac{\epsilon}{2j-1}\left(
\frac{a-b}{j+1/2}+\frac{a+b}{j-1/2}\right)  -\frac{2}{2j-1}\right]  ,
\label{largejchar}%
\end{equation}

Applying the same approach to the scattering

Our numerical analysis shows that the energy dependence of the relaxation time
crosses over to an approximate constant at high energy, and position of the
crossing-over point depends on the ratio $a/b$ (see fig. 1). The Boltzmannian
conductivity is determined by the formula:%
\begin{equation}
\sigma=\frac{e^{2}}{h}\left(  E_{F}\tau_{tr}/\hbar\right)  ,
\label{conductgeneral}%
\end{equation}
where $\tau_{tr}$ is determined as follows
\begin{equation}
\tau_{tr}^{-1}=N_{I}\Sigma_{tr}v_{F}. \label{timecross}%
\end{equation}

The mobility can be determined as the ratio:%

\begin{equation}
\mu=\frac{\sigma}{en}, \label{mobilgeneral}%
\end{equation}
where the carrier density at low temperature is determined as follows%
\begin{equation}
n=N/S=\frac{1}{2\pi}\left(  \frac{E_{F}}{\hbar v_{F}}\right)  ^{2}
\label{fermi}%
\end{equation}

Using (\ref{time2}), (\ref{conductgeneral}), and (\ref{timecross}) we conclude
that the conductivity tends to a constant value in the limit $E_{F}%
\rightarrow0.$This limit must be considered taking into account the fact tha
the accepted here Boltzmann kinetics in invalid in the vicinity of the point
$E_{F}=0.$ In the opposite limit of large $E_{F},$ we conclude using
(\ref{crosslargeenergy}) and (\ref{timecross}) instead of (\ref{time2}) that
conductivity increases linearly with the Fermi energy in this energy region.
The mobility behaves at small and large Fermi energy respectively as
$E_{F}^{-2}$ and $E_{F}^{-1}.$

\end{document}